\documentclass[aps,prd,twocolumn,showpacs,groupedaddress]{revtex4}
\usepackage{graphicx}
\usepackage{amsfonts}
\usepackage{amssymb}
\usepackage{amsmath}

\begin{document}

\title{
  Lattice QCD analysis for Faddeev-Popov eigenmodes \\
  in terms of gluonic momentum components in the Coulomb gauge
}

\author{Takumi~Iritani}
  \email{iritani@ruby.scphys.kyoto-u.ac.jp}
  \affiliation{Department of Physics, Graduate School of Science,
  Kyoto University, \\
  Kitashirakawa-oiwake, Sakyo, Kyoto 606-8502, Japan}
\author{Hideo~Suganuma}
  \email{suganuma@ruby.scphys.kyoto-u.ac.jp}
  \affiliation{Department of Physics, Graduate School of Science,
  Kyoto University, \\
  Kitashirakawa-oiwake, Sakyo, Kyoto 606-8502, Japan}
\date{\today}

\begin{abstract}
  We analyze the relation between Faddeev-Popov eigenmodes
  and gluon-momentum components in the Coulomb gauge using SU(3) lattice QCD.
  In the Coulomb gauge, the color-Coulomb energy is largely enhanced 
  by near-zero Faddeev-Popov eigenmodes, which would lead to the confining potential.
  By the ultraviolet-momentum gluon cut,
  the color-Coulomb energy and the Faddeev-Popov spectrum are almost unchanged.
  In contrast to the ultraviolet insensitivity, 
  the color-Coulomb energy and
  the Faddeev-Popov eigenmodes drastically change by infrared-momentum gluon cut.
  Without infrared gluons,
  the color-Coulomb energy tends to become non-confining,
  and near-zero Faddeev-Popov eigenmodes vanish.
  We also investigate the full FP eigenmodes, 
  and find that infrared gluons widely influence 
  both high and low Faddeev-Popov eigenmodes.
\end{abstract}


\maketitle
\section{Introduction}

  The Coulomb gauge is one of the most popular gauges in 
  Quantum Chromodynamics (QCD).
  In addition to the convenient choice for calculations,
  the Coulomb-gauge QCD is also interesting from the theoretical point of view,
  especially in the context of the canonical quantization \cite{ItzyksonZuber}.
  In addition, color confinement has been investigated in the Coulomb gauge
  in terms of the color-Coulomb interaction
  \cite{Gribov:1978,Zwanziger:2003,
  Greensite:2003,Cucchieri:2002,Cucchieri:2003,Cucchieri:2006,
  Langfeld:2004,
  Greensite:2004,Greensite:2005,
  Nakamura:2005,Nakagawa:2007,Nakagawa:2010,
  Feuchter:2004,Voigt:2008,Szczepaniak:2006}.
  As a new picture of hadrons, 
  the gluon-chain picture also stems from the Coulomb-gauge QCD
  \cite{Greensite:2002,tHooft:2003,Greensite:2009}.

  In the Coulomb-gauge QCD, 
  the Faddeev-Popov (FP) operator is the important key quantity.
  Actually, the color-Coulomb energy is enhanced by the near-zero FP eigenmodes, 
  which would lead to the confining force between color charges.
  This confinement scenario is known as the Gribov-Zwanziger scenario
  \cite{Gribov:1978,Zwanziger:2003}.
  From lattice-QCD numerical calculations,
  the color-Coulomb energy certainly
  gives a linear interquark potential
  \cite{Cucchieri:2002, Cucchieri:2003, Cucchieri:2006,
  Langfeld:2004, Greensite:2003, Greensite:2004, Greensite:2005,
  Nakamura:2005,Nakagawa:2007,Nakagawa:2010,Voigt:2008}.

  The confinement is also characterized by the infinite self-energy
  of an isolated color-charge.
  In the Coulomb gauge, 
  the color-Coulomb self-energy is expressed
  in terms of the FP eigenmodes,
  and its divergence is originated 
  from near-zero FP eigenmodes,
  as was shown in lattice QCD calculations 
  \cite{Greensite:2005,Nakagawa:2007}.

  Then, it is meaningful to investigate the FP eigenmodes
  for the understanding of non-perturbative QCD
  \cite{Greensite:2005,Greensite:2005b,Nakagawa:2007,Maas:2006}.
  In this paper, we investigate the FP eigenmodes 
  in terms of the gluon-momentum components.
  Since the QCD interactions are mediated by gluons,
  the FP eigenmodes should reflect the properties of the gluon.
  To analyze the correspondence between the gluon field and the FP eigenmodes,
  we decompose the link-variables in terms of momentum components,
  and remove infrared or ultraviolet momentum modes in lattice QCD \cite{Yamamoto:2008}.


  The organization of this paper is as follows.
  In Sec.II, we review the color-Coulomb energy
  and properties of the FP eigenmodes in the Coulomb gauge.
  We also introduce the gluon-momentum cut method in lattice QCD.
  In Sec.III, we show the lattice QCD results for the color-Coulomb energy
  and the FP eigenmodes with the IR/UV-momentum gluon cut.
  Section IV will be devoted to summary and discussions.

\section{Formalism}
  In this section, we review the color-Coulomb energy
  and the FP operator properties in the Coulomb gauge
  \cite{Greensite:2003,Greensite:2004,Greensite:2005}.
  We also briefly introduce the gluon-momentum cut method in lattice QCD.
  Since we discuss the Coulomb-gauge QCD at a fixed timeslice $t$,
  we will omit the time coordinate $t$ in this paper.
 
\subsection{Color-Coulomb energy and confinement scenario in the Coulomb gauge}
  The Coulomb gauge is one of the most popular gauges
  in both analytic framework and lattice QCD calculation.
  The definition of the Coulomb gauge is given by
  \begin{equation}
    \partial_i A_i = 0,
  \end{equation}
  where the gluon fields $A_\mu(\vec{x}) \equiv A_\mu^a(\vec{x})T^a \in \mathfrak{su}(N_c)$
  with generator $T^a$ ($a=1,2,\dots, N_c^2-1$).
  The Coulomb gauge is also defined by
  the minimization of the global quantity
  \begin{equation}
    R_{\rm Coul} \equiv \int d^3 \vec{x} \ \mathrm{Tr} \{ A_i(\vec{x}) A_i(\vec{x}) \},
    \label{eqCoulombGlobal}
  \end{equation}
  by the gauge transformation.
  The minimization of the quantity $R_{\rm Coul}$
  means that the spatial gauge-field fluctuations
  are maximally suppressed in the Coulomb gauge.
 
  These gauge-fixing conditions resemble the Landau-gauge condition,
  although the Lorentz covariance is partially broken.
  In the Coulomb gauge, gluon fields are decomposed into the canonical 
  variable $\vec{A}$ and the potential $A_0$ \cite{Iritani:2011}.

  One of the advantages in the Coulomb gauge
  is the compatibility with the canonical quantization
  \cite{ItzyksonZuber}.
  In the Coulomb gauge, 
  the QCD  Hamiltonian is expressed 
  with color electric and magnetic fields, $\vec{E}^a$ and $\vec{B}^a$,
  as
  \begin{eqnarray}
    H &=& \frac{1}{2} \int d^3\vec{x} \left( \vec{E}^a \cdot \vec{E}^a
    + \vec{B}^a \cdot \vec{B}^a \right) \nonumber \\ 
    &&\quad + \frac{1}{2} \int d^3\vec{x} d^3\vec{y} 
    \rho^a(\vec{x}) K^{ab}(\vec{x},\vec{y}) \rho^b(\vec{y}),
    \label{eq:QCDHamiltonian}
  \end{eqnarray}
  where 
  $\rho^a(\vec{x})$ is the color-charge density, and $K^{ab}(\vec{x},\vec{y})$
  the instantaneous Coulomb propagator. 
  $K^{ab}(\vec{x},\vec{y})$ is defined as
  \begin{equation}
    K^{ab}(\vec{x},\vec{y}) 
    = [M_{\rm FP}^{-1}(-\nabla^2)M_{\rm FP}^{-1}]_{\vec{x}\vec{y}}^{ab},
  \end{equation}
  using the FP operator
  \begin{equation}
    M_{\rm FP}^{ac} = - \nabla^2 \delta^{ac} - \varepsilon^{abc}A_i^b \partial_i,
    \label{eq:FPopContinuum}
  \end{equation}
  with $\nabla^2 = \partial_i^2$.
  The second term of the Hamiltonian (\ref{eq:QCDHamiltonian}) 
  corresponds to the color-Coulomb energy,
  which is expressed as
  \begin{equation}
    V_{\rm Coul}(R) = 
    -g^2 \frac{C_F}{N_c^2-1} \langle 
    \left[ M_{\rm FP}^{-1}(-\nabla^2) M_{\rm FP}^{-1} \right]_{\vec{x}\vec{y}}^{aa}
    \rangle
    \label{eqVcoul}
  \end{equation}
  with $R = |\vec{x} - \vec{y}|$, 
  the quadratic Casimir $C_F = \frac{N_c^2-1}{2N_c}$ of
  the fundamental representation, and the coupling constant $g$ 
  \cite{Greensite:2003,Greensite:2004,Greensite:2005}.

  In the Abelian gauge theory,
  the FP operator becomes the Laplacian,
  and the instantaneous Coulomb propagator is inverse of the Laplacian. 
  Thus, the Coulomb energy becomes familiar Coulomb potential form $V(R) \propto 1/R$.
  In the non-Abelian gauge theory,
  the FP operator has non-trivial zero-modes, which form the Gribov horizon
  \cite{Gribov:1978}.
  In the neighborhood of the Gribov horizon,
  the color-Coulomb energy is largely enhanced, 
  which is conjectured to contribute to the confining force.
  This confinement picture is known as
  the ``Gribov-Zwanziger scenario'' \cite{Gribov:1978, Zwanziger:2003}.

  Instead of the direct definition of the color-Coulomb energy 
  $V_{\rm Coul}(R)$ in Eq.(\ref{eqVcoul}),
  we consider the correlator of the time-like Wilson lines as
  \begin{equation}
    G(R,T) \equiv \frac{1}{N_c} \langle  \mathrm{Tr}
    \left[ L(\vec{x},T) L^\dagger(\vec{y},T)\right] \rangle,
  \end{equation}
  with $R = | \vec{x} - \vec{y}|$ and the time-like Wilson line
  \begin{equation}
    L(\vec{x},T) \equiv P \exp\left\{i\int_0^T dt A_4(\vec{x},t)\right\}.
  \end{equation}
  As for the relation to the color-Coulomb energy, one finds
  \cite{Greensite:2003,Greensite:1986}
  \begin{equation}
    - \left. \frac{d}{dT} \ln  G(R,T) \right|_{T\rightarrow 0}
    = V_{\rm Coul}(R) + \mathrm{const.}
    \label{eqVcoulDef2}
  \end{equation}

  In lattice QCD, 
  the QCD action is constructed from the link-variable $U_\mu(\vec{x})$, 
  which is defined as $U_\mu(\vec{x}) = e^{iagA_\mu(\vec{x})}$ 
  with the lattice spacing $a$ and the gauge coupling constant $g$ \cite{Rothe}.
  The Coulomb-gauge fixing condition is 
  expressed in terms of the link-variable,
  and is given by the maximization of
  \begin{equation}
    R[U] \equiv \sum_{\vec{x}} \sum_{i=1}^3 \mathrm{Re} \ \mathrm{Tr} \ U_i(\vec{x}),
    \label{eq:R_Coul}
  \end{equation}
  by the gauge transformation
  \begin{equation}
    U_i(\vec{x})  \rightarrow 
    \Omega(\vec{x}) U_i(\vec{x}) \Omega^\dagger(\vec{x}+\hat{\imath}),
  \end{equation}
  with $\Omega(\vec{x})\in \mathrm{SU}(N_c)$.
  In the continuum limit $a \rightarrow 0$, this condition results in 
  the minimization of Eq.(\ref{eqCoulombGlobal}).
  Using the time-like Wilson line correlator $G(R,T)$ on lattice,
  we define $V(R,T)$ as
  \begin{equation}
    V(R,T) \equiv\frac{1}{a} \ln \left[ \frac{G(R,T)}{G(R,T+a)} \right].
  \end{equation}
  Especially, at $T = 0$, we call 
  \begin{equation}
    V_{\rm inst}(R) \equiv - \frac{1}{a} \ln 
    \langle \mathrm{Tr} \{ U_4(\vec{x}) U_4^\dagger(\vec{y}) \} \rangle
    \label{eq:VcoulDef}
  \end{equation}
  as ``instantaneous potential'' \cite{Iritani:2011},
  which is considered to be closely related to the color-Coulomb energy 
  in Eq.(\ref{eqVcoul}).
  Actually, in the continuum limit, the instantaneous potential would coincide
  with the color-Coulomb energy as in Eq.(\ref{eqVcoulDef2}) \cite{Greensite:2003,Greensite:2004}.
  The lattice QCD calculations show 
  that the instantaneous potential gives a linear rising potential
  \cite{Greensite:2003,Greensite:2004},
  and satisfies the Casimir scaling \cite{Nakamura:2005}
  similar to the physical interquark potential \cite{Bali:2001}.
  However, the slope of the potential is $2 \sim 3$ times
  larger than physical string tension $\sigma_{\rm phys} \simeq 0.89$GeV/fm
  \cite{Greensite:2003,Greensite:2004}.

  The color-Coulomb energy is directly calculated 
  based on Eq.(\ref{eqVcoul}) in both SU(2) \cite{Cucchieri:2003, Langfeld:2004} 
  and SU(3) \cite{Nakagawa:2010} lattice QCD.
  These lattice QCD calculations also indicate the overconfining potential
  with a larger string tension.

  Actually, the color-Coulomb energy gives 
  an upper bound on the static interquark potential $V_{\rm phys}(R)$,
  \begin{equation}
    V_{\rm phys}(R) \leq V_{\rm Coul}(R),
  \end{equation}
  which was shown by Zwanziger \cite{Zwanziger:2003}.
  This large color-Coulomb energy
  is considered as the overconfining state,
  and the gluon-chain picture is proposed for
  the true ground state of the quark-antiquark system
  in the Coulomb gauge \cite{Greensite:2003,Greensite:2002,Greensite:2009}.

  In spite of overconfining, 
  the color-Coulomb energy is expected to relate
  to the confinement at least at zero temperature.
  Therefore, we concentrate on the color-Coulomb energy
  properties in this paper.

\subsection{Faddeev-Popov eigenmodes and color-Coulomb energy}
  In the Coulomb-gauge lattice QCD, 
  the FP operator is given by
  \begin{eqnarray}
    &&M_{\rm FP}(a,\vec{x};b,\vec{y}) \equiv \nonumber \\
    &&\quad \sum_{i=1}^3 \mathrm{Re} \mathrm{Tr} \Big[ \left\{ T^a,T^b \right\} 
    \left( U_i(\vec{x}) + U_i(\vec{x}-\hat{\imath}) \right)
    \delta_{\vec{x},\vec{y}} \nonumber \\
    &&\quad -2 T^b T^a U_i(\vec{x}) \delta_{\vec{x}+\hat{\imath},\vec{y}}
    -2 T^a T^b U_i(\vec{x}-\hat{\imath}) \delta_{\vec{x}-\hat{\imath},\vec{y}}\Big]
  \end{eqnarray}
  using the generators $T^a$ and link-variable $U_\mu(\vec{x})$.
  On $L^3 \times L_t$ lattice,
  the total number of FP eigenmodes is $V_3 \times (N_c^2-1)$
  with the spatial volume $V_3 = L^3$.
  The FP operator $M_{\rm FP}(a,\vec{x};b,\vec{y})$ has 
  trivial $(N_c^2-1)$ zero-modes
  \begin{equation}
    \psi_n^{a}(\vec{x}) \equiv \frac{1}{\sqrt{V_3}} \delta_{an},
    \label{eq:TrivialZero}
  \end{equation}
  with $a = 1,2,\dots,(N_c^2-1)$.

  In free-field and QED cases,
  the FP operator becomes the Laplacian,
  and the eigenvalues are expressed by the three-dimensional momentum
  $p_i$.
  In the lattice theory, 
  the eigenvalue is given by
  \begin{equation}
    \lambda = \sum_{i=1}^3 
    \left( \frac{2}{a} \sin\left( \frac{p_i a}{2} \right) \right)^2,
    \label{eq:FPeigenFree}
  \end{equation}
  with $p_i \in (-\pi/a,\pi/a]$.
  In the continuum limit, it becomes
  \begin{equation}
    \lambda^{(\rm cont.)}  = \sum_{i=1}^3 p_i^2  = \vec{p}^{\ 2}.
    \label{eq:FPeigenFreeCont}
  \end{equation}
 
  In terms of the FP eigenfunction,
  the relation between the color-Coulomb energy
  and the Gribov horizon becomes clear.
  Considering the FP eigenstate $|\lambda_n \rangle$ which satisfies
  \begin{equation}
    M_{\rm FP}|\lambda_n \rangle = \lambda_n | \lambda_n \rangle,
  \end{equation}
  with eigenvalue $\lambda_n \in \mathbf{R}$,
  and the FP eigenfunction is given by
  \begin{equation}
    \psi_n^a(\vec{x}) \equiv \langle \vec{x},a | \lambda_n \rangle,
  \end{equation}
  where $a$ is the color index.
  Using the FP eigenmodes,
  the color-Coulomb energy in Eq.(\ref{eqVcoul}) is expressed as
  \begin{equation}
    V_{\rm Coul}(R) = - g^2 \frac{C_F}{N_c^2-1}
    \sum_{n,m} \psi_n^a(\vec{x})\psi_m^{a \ast}(\vec{y})
    \frac{\langle \lambda_n | - \nabla^2 | \lambda_m \rangle}{
    \lambda_n \lambda_m}
    \label{eq:VcoulFPmode}
  \end{equation}
  with $R = |\vec{x} - \vec{y}|$ \cite{Greensite:2005}. 
  Equation (\ref{eq:VcoulFPmode})
  indicates that low-lying FP eigenmodes
  would give dominant contribution to the color-Coulomb energy.
  From lattice QCD calculations,
  the color-Coulomb energy is brought by only small number of
  low-lying FP eigenmodes \cite{Nakagawa:2010}.

  The confinement is also investigated in terms of the color-Coulomb self-energy,
  since an isolated color-charge has infinite energy in the infrared in QCD.
  Using the FP eigenmodes, the color-Coulomb self-energy 
  \cite{Greensite:2005}
  is expressed as
  \begin{eqnarray}
    \mathcal{E}_F &=& \frac{g^2C_F}{N_c^2 - 1}
    \langle K^{aa}(\vec{x},\vec{x}) \rangle \nonumber \\
    &=& \frac{g^2C_F}{N_c^2-1}\frac{1}{V_3} \sum_n 
    \frac{\langle \lambda_n | - \nabla^2 | \lambda_n \rangle}{\lambda_n^2} \nonumber \\
    &=& \frac{g^2C_F}{N_c^2-1}\frac{1}{V_3} \sum_n \frac{F(\lambda_n)}{\lambda_n^2}.
  \end{eqnarray}
  $F(\lambda_n)$ is the diagonal-matrix element of the Laplacian operator:
  \begin{eqnarray}
    F(\lambda_n)
    &\equiv& \langle \lambda_n | - \nabla^2 | \lambda_n \rangle \nonumber \nonumber \\
    &=& \sum_{\vec{x}, \vec{y}} \sum_{i=1}^3 
    \psi_n^{a \ast}(\vec{x}) \nonumber \\
    &&\qquad \times [ 2\delta_{\vec{x},\vec{y}} 
    - \delta_{\vec{x}+\hat{\imath},\vec{y}}
    - \delta_{\vec{x}-\hat{\imath},\vec{y}} ] \psi_n^a(\vec{y}).
  \end{eqnarray}
  Here, we define the eigenmode density $\rho(\lambda)$ as
  \begin{equation}
    \rho(\lambda) \equiv 
    \frac{1}{(N_c^2-1)V_3} \frac{1}{\Delta \lambda}
    N(\lambda,\lambda+\Delta\lambda),
  \end{equation}
  where $N(\lambda,\lambda+\Delta\lambda)$ is the number of eigenvalues
  in $[\lambda, \lambda+\Delta\lambda]$.
  In the infinite-volume limit, color-Coulomb self-energy $\mathcal{E}_F$ 
  \cite{Greensite:2005} is expressed as
  \begin{equation}
    \mathcal{E}_F = g^2C_F \int_0^{\lambda_{\rm max}}
    \frac{d\lambda}{\lambda^2}\rho(\lambda)F(\lambda),
  \end{equation}
  with the FP-eigenmode density $\rho(\lambda)$, and the UV cutoff $\lambda_{\rm max}$.
  Therefore, if the criterion
  \begin{equation}
    \lim_{\lambda\rightarrow 0} \frac{\rho(\lambda)F(\lambda)}{\lambda} > 0
    \label{eqSelfEnergyDivergence}
  \end{equation}
  is satisfied, 
  color-Coulomb self-energy $\mathcal{E}_F$ diverges
  in the infrared limit \cite{Greensite:2005}.
  This criterion indicates the importance
  of the near-zero FP modes and matrix elements $F(\lambda)$ for confinement,
  which is similar to the Dirac zero-mode and chiral symmetry breaking
  in the Banks-Casher relation \cite{BanksCasher}.

  In free-field and QED cases, 
  the FP eigenvalue is given by Eq.(\ref{eq:FPeigenFreeCont}),
  and then $\rho(\lambda)$ and $F(\lambda)$ behave as
  \begin{equation}
    \rho(\lambda) \sim \lambda^{1/2}, F(\lambda) \sim \lambda.
  \end{equation}
  Therefore, the confinement criterion is not satisfied as
  \begin{equation}
    \lim_{\lambda \rightarrow 0} \frac{\rho(\lambda)F(\lambda)}{\lambda} = 0,
  \end{equation}
  and the Coulomb self-energy is infrared finite.

  Since the color-charge is confined in QCD, 
  color-Coulomb self-energy is expected to diverge at infrared limit.
  From SU(2) lattice-QCD analysis,
  the FP spectrum $\rho(\lambda)$ and the matrix element $F(\lambda)$
  approximately behave as
  \begin{equation}
    \rho(\lambda) \sim \lambda^{0.25}, \quad
    F(\lambda) \sim \lambda^{0.38},
  \end{equation}
  near the horizon $\lambda \sim 0$ \cite{Greensite:2005}.
  Then the divergence criterion is satisfied as
  \begin{equation}
    \lim_{\lambda \rightarrow 0}
    \frac{\rho(\lambda)F(\lambda)}{\lambda} \sim 
    \lim_{\lambda\rightarrow 0}\lambda^{-0.37} = \infty.
  \end{equation}
  Therefore, the color-Coulomb self-energy actually diverges 
  in the infrared limit and the color-charge is confined
  in the non-Abelian case.

\subsection{Gluon-momentum cut method}
  In this subsection, we introduce
  the formalism to remove gluon-momentum modes
  in lattice QCD, which proposed in Ref.\cite{Yamamoto:2008}.
  Here, we use the Coulomb gauge,
  and carry out the three-dimensional Fourier transformation
  with respect to the spatial coordinate $\vec{x}$ at fixed timeslice $t$.

  The procedure is as follows.

  1. We generate the link-variable $U_\mu(x) \in \mathrm{SU}(N_c)$
  on $L^3 \times L_t$ lattice 
  with lattice spacing $a$,
  and fix the Coulomb gauge.

  2. We carry out the discrete Fourier transformation.
  The momentum-space link-variable is given by
  \begin{equation}
    \tilde{U}_\mu(\vec{p}) 
    \equiv \frac{1}{V_3}
    \sum_{\vec{x}} U_\mu(\vec{x}) \exp (i \sum_{i=1}^3 p_i x_i ),
  \end{equation}
  with the spatial volume $V_3 = L^3$,
  and momentum $p_i \in (-\pi/a,\pi/a]$.
  In the momentum space, the lattice spacing $a_p$ is given by
  \begin{equation}
    a_p \equiv \frac{2\pi}{L a}.
  \end{equation}

  3. We introduce infrared and ultraviolet ``cut''
  $\Lambda_{\rm IR(UV)} \in (-\pi/a,\pi/a]$,
  and replace the link-variable by 
  the free-field link-variable as
  \begin{equation}
    \tilde{U}_\mu^\Lambda(\vec{p}) 
    \equiv
    \begin{cases}
      \tilde{U}_\mu(\vec{p}) 
        & \Lambda_{\rm IR}^2 \leq p_i^2 \leq \Lambda_{\rm UV}^2  \\
      \tilde{U}_\mu^{\rm free}(\vec{p}) 
        &  p_i^2 < \Lambda_{\rm IR}^2 \ \text{or} \ \Lambda_{\rm UV}^2 < p_i^2.
    \end{cases}
  \end{equation}
  The Fourier transformation of 
  the free-field link-variable
  $U_\mu(\vec{x}) = \mathbf{1}$ is
  \begin{equation}
    \tilde{U}_\mu^{\rm free}(\vec{p}) \equiv
    \frac{1}{V_3} \sum_{\vec{x}} \mathbf{1}
    \cdot \exp ( i\sum_{i=1}^3 p_i x_i )
    = \delta_{\vec{p}\vec{0}} \mathbf{1}.
  \end{equation}

  4. In order to return to the coordinate space variable,
  we carry out the inverse Fourier transformation of $\tilde{U}_\mu^\Lambda(\vec{p})$ as
  \begin{equation}
    U_\mu^{\prime \Lambda}(\vec{x}) \equiv \sum_{\vec{p}} 
    \tilde{U}_\mu^\Lambda(\vec{p}) \exp (  - i\sum_{i=1}^3 p_i x_i ).
  \end{equation}
  Since this link-variable $U_\mu^{\prime \Lambda}(\vec{x})$
  is not SU($N_c$) matrix,
  we project it onto SU($N_c$) matrix $U_\mu^\Lambda(x)$ by maximizing
  \begin{equation}
    \mathrm{Re} \ \mathrm{Tr} 
    [U_\mu^{\prime \Lambda}(\vec{x})U_\mu^{\Lambda \dagger}(\vec{x})],
  \end{equation}
  which is often used in lattice QCD algorithm.
  Finally, we obtain the momentum projected link-variable
  $U_\mu^\Lambda(\vec{x}) \in \mathrm{SU}(N_c)$.

\section{Lattice QCD calculation}
  In this section, we analyze the instantaneous potential,
  the FP eigenmodes, and the color-Coulomb energy,
  in terms of the gluonic momentum 
  using SU(3) lattice Monte Carlo calculations.
  We mainly use $16^4$ lattice at $\beta \equiv 2N_c/g^2 = 5.8$,
  which corresponds to the lattice spacing
  $a \simeq 0.15$fm \cite{Suganuma:2001},
  and $a_p \equiv 2\pi/La \simeq 0.50$GeV in the momentum space.
  We use the jack-knife method for the estimating of the statistical error.

\subsection{Instantaneous potential with IR/UV gluon cut}
\begin{figure}
  \centering
  \includegraphics[width=0.5\textwidth,clip]{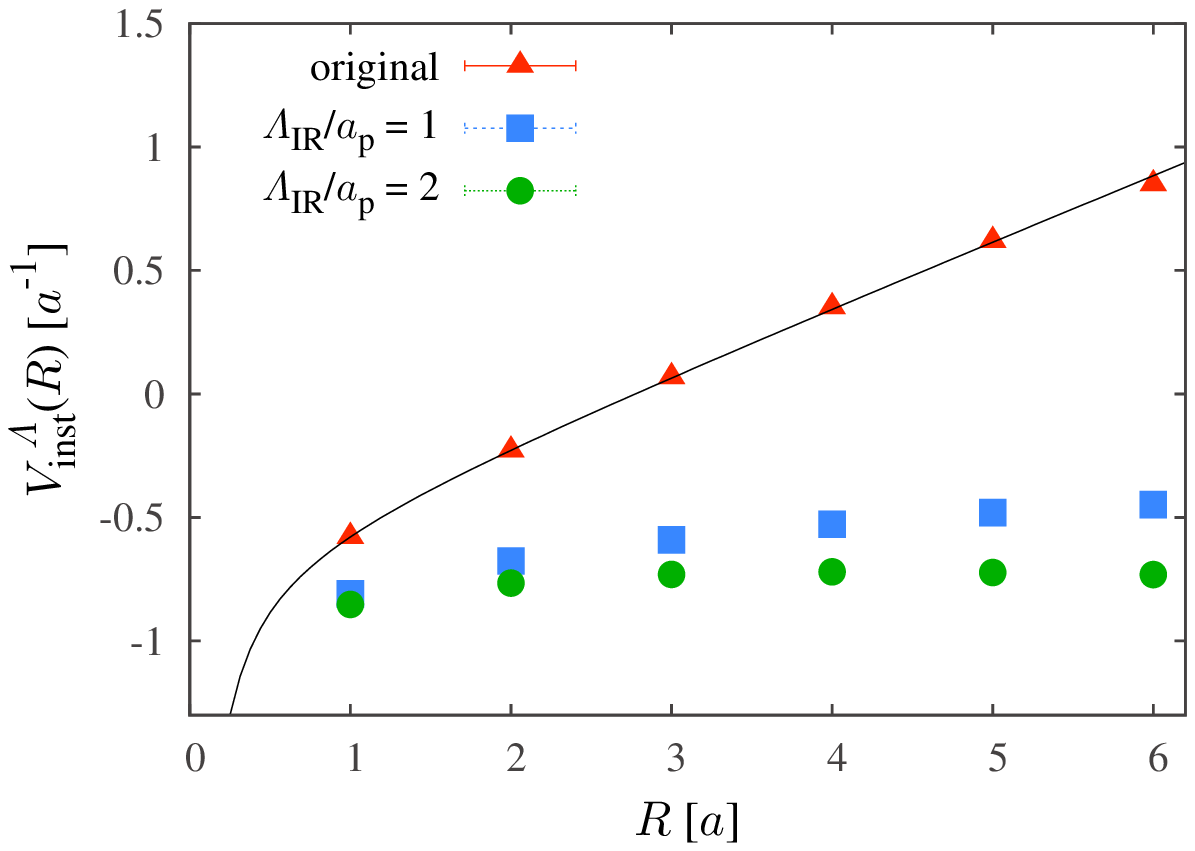}
  \includegraphics[width=0.5\textwidth,clip]{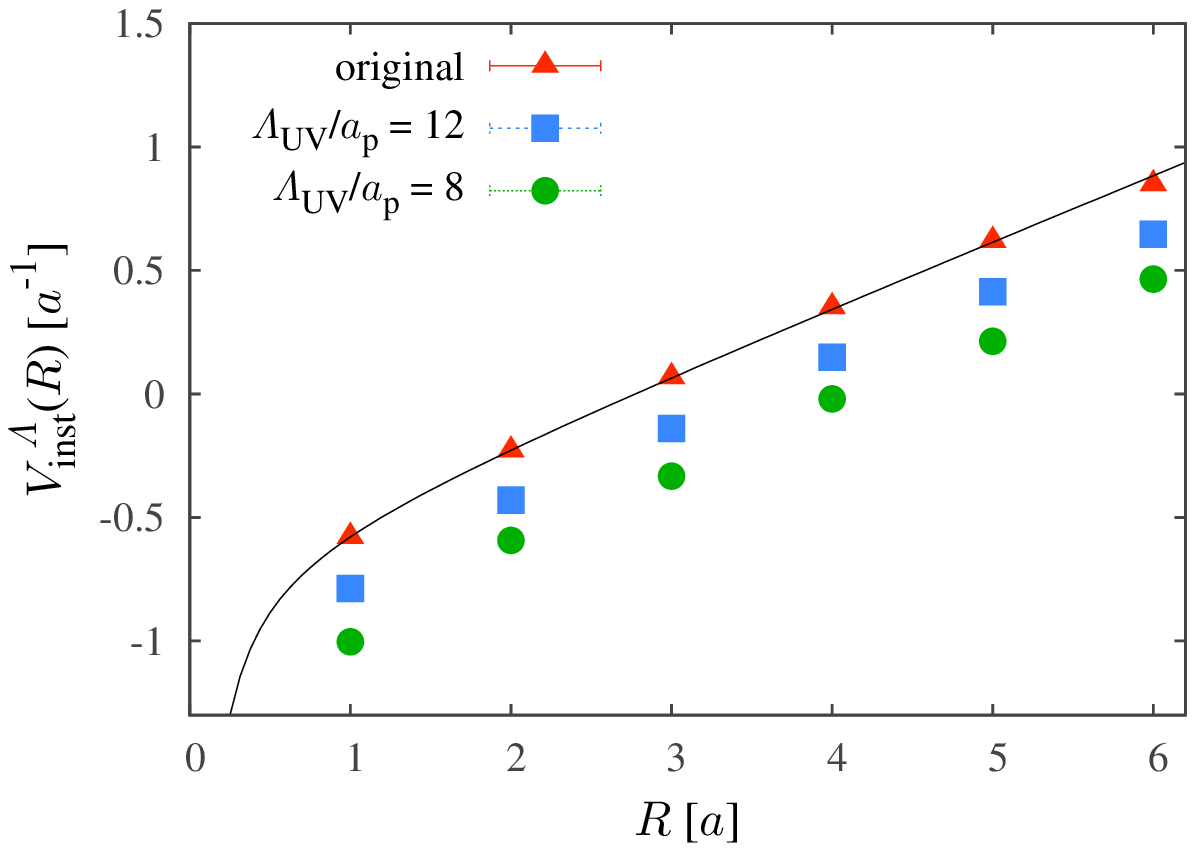}
  \caption{ \label{figColorCoulomb}
  The instantaneous potential $V_{\rm inst}^\Lambda(R)$ 
  with the IR/UV-momentum gluon cut
  for $16^4$ lattice with $a \simeq 0.15$fm, i.e., $a_p \equiv 2\pi/La \simeq 0.50$GeV.
  The original (no momentum cut) instantaneous potential
  is added with the fitting curve of
  Coulomb plus linear form.
  The statistical error is small and the error bars are hidden in the symbols.
  (a) IR-momentum cut with $\Lambda_{\rm IR}/a_p = 1$ and $2$.
  (b) UV-momentum cut with $\Lambda_{\rm UV}/a_p = 12$ and $8$.
  An irrelevant constant is shifted.}
\end{figure}

  First, we analyze the instantaneous potential
  with the IR/UV-momentum gluon cut.
  Using $U_\mu^\Lambda(\vec{x})$,
  we define the instantaneous potential
  with gluon-momentum cut as
  \begin{equation}
    V_{\rm inst}^\Lambda(R) \equiv - \frac{1}{a}
    \ln \langle \mathrm{Tr}\{U_4^\Lambda(\vec{x})U_4^{\Lambda\dagger}(\vec{y})\}\rangle,
  \end{equation}
  with $R = | \vec{x} - \vec{y}|$.
  Figure \ref{figColorCoulomb}(a)
  and (b)
  are $V_{\rm inst}^\Lambda(R)$
  with the IR/UV-momentum cut, respectively.
  We also show original (no momentum cut) instantaneous potential,
  and fit result using Coulomb plus linear form.
  The best fit value of the slope is
  $\sigma_{\rm Coul}/\sigma_{\rm phys} \simeq 2.6$.

  As shown in Fig.\ref{figColorCoulomb}(a),
  the instantaneous potential becomes non-confining with IR-momentum cut.
  The instantaneous potential changes drastically even for the smallest IR-cut.
  In contrast to the IR-cut,
  the instantaneous potential is almost unchanged by the UV-cut
  as shown in Fig.\ref{figColorCoulomb}(b).

\subsection{FP eigenmodes with IR/UV gluon cut}
  Next, we analyze the FP eigenmodes with the IR/UV-momentum gluon cut.
  The instantaneous potential becomes non-confining with the IR-cut.
  Similar to the color-Coulomb energy,
  the instantaneous potential would be closely related to the FP eigenmodes,
  we expect that the FP eigenmodes are largely changed with the IR-cut.

\subsubsection{Low-lying FP eigenmodes}
  First, we evaluate the low-lying 250 FP eigenmodes 
  using ARPACK \cite{ARPACK}.
  In this case, the total number of FP eigenmodes is 
  $V_3 \times (N_c^2-1) = 16^3 \times 8 = 32768$.

\begin{figure}[h]
  \centering
  \includegraphics[width=0.5\textwidth,clip]{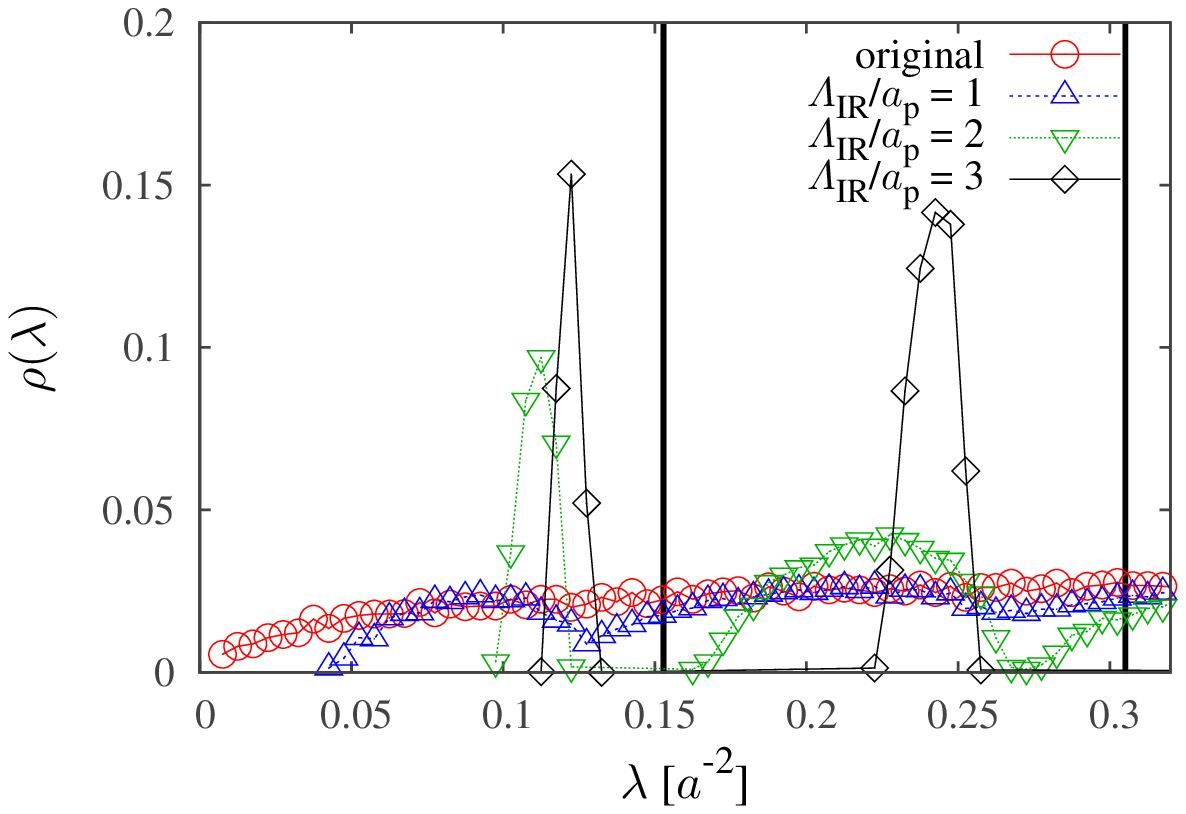}
  \includegraphics[width=0.5\textwidth,clip]{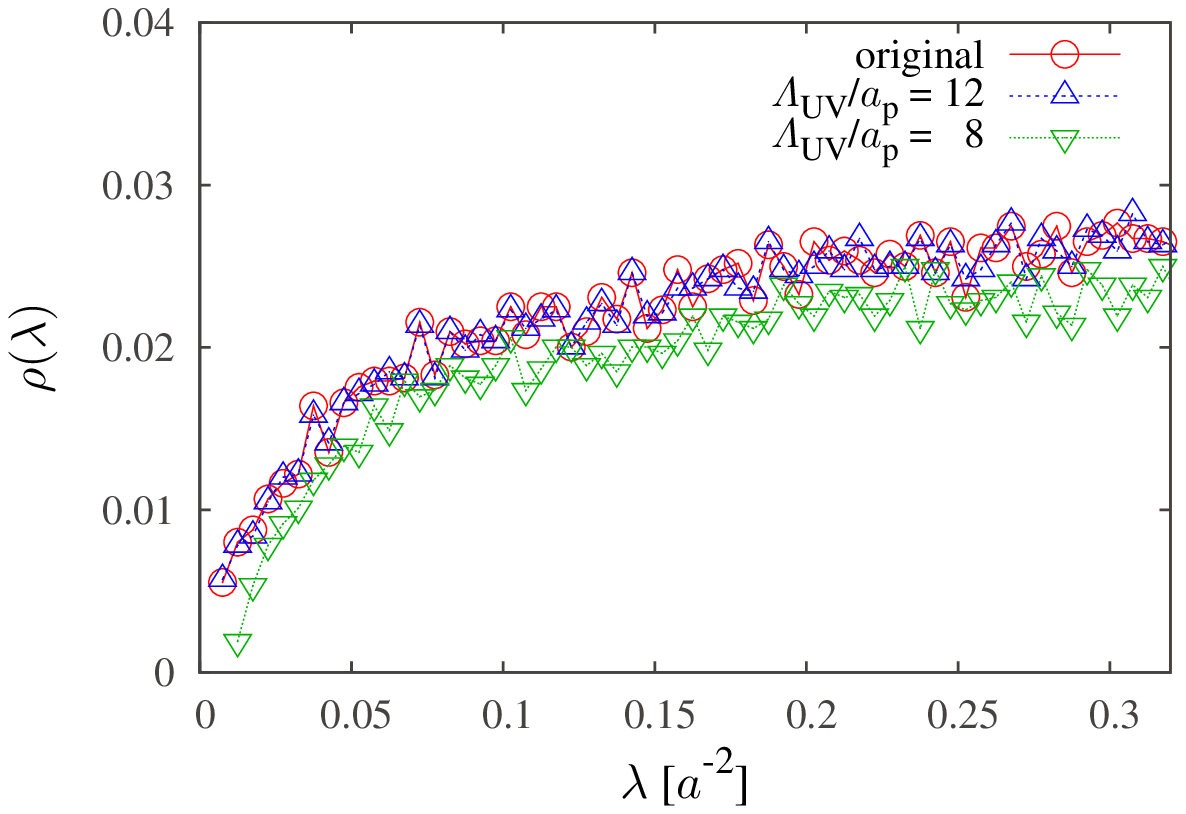}
  \caption{ \label{figFPspectrumLow}
    Low-lying 250 FP spectrum $\rho(\lambda)$ with the IR/UV momentum cut and original spectrum
    for $16^4$ lattice with $a \simeq 0.15$fm, i.e., $a_p \equiv 2\pi/La \simeq 0.50$GeV.
    The binwidth is taken as $\Delta \lambda = 0.005a^{-2}$.
    (a) IR-momentum cut with $\Lambda_{\rm IR}/a_p = 1,2$, and $3$.
    The vertical bars denote the non-zero free-field spectrum.
    (b) UV-momentum cut with $\Lambda_{\rm UV}/a_p = 12$ and $8$.
  }
\end{figure}

  Figure \ref{figFPspectrumLow}(a) and (b) are
  the low-lying FP spectrum $\rho(\lambda)$ with
  the IR/UV-momentum gluon cut, respectively.
  In both figures, we have added the original FP spectrum
  for comparison, and we have omitted the trivial 8 zero-modes
  in Eq.(\ref{eq:TrivialZero}),
  which always remain even after momentum cut procedure.
  In Fig.\ref{figFPspectrumLow}(a),
  the vertical bars denote the free-field eigenvalues
  $\lambda = 4\sin^2(\pi/16)a^{-2} \simeq 0.15a^{-2}$
  and $\lambda = 8\sin^2(\pi/16)a^{-2} \simeq 0.30a^{-2}$.

  As shown in Fig.\ref{figFPspectrumLow}(a), 
  the IR-cut FP spectrum is drastically changed.
  The near-zero FP modes vanish, and the spectrum changes into multi-peak structure
  from original smooth one.
  By increasing the IR-cut $\Lambda_{\rm IR}$, 
  the peaks of the spectrum become sharper,
  and the FP spectrum tends to converge into $\delta$-functional peaks 
  in the free-field limit as Eq.(\ref{eq:FPeigenFree}).
  On the other hand, the FP spectrum is almost unchanged for the UV-cut.
  This UV-cut insensitivity is the same as the instantaneous potential
  as shown in Fig.\ref{figColorCoulomb}(b).

\subsubsection{Matrix element of the Laplacian operator}
  Next, we analyze the matrix element of the Laplacian operator,
  i.e., $\langle \lambda_1 | - \nabla^2 | \lambda_2 \rangle$,
  which is important for both color-Coulomb energy in Eq.(\ref{eq:VcoulFPmode}) 
  and self-energy divergence condition (\ref{eqSelfEnergyDivergence}).
  Here, we mainly discuss the diagonal component
  $F(\lambda) \equiv \langle \lambda | - \nabla^2 | \lambda \rangle$,
  since the off-diagonal elements
  $\langle \lambda_1 | - \nabla^2 | \lambda_2 \rangle$
  are found to be almost the zero from lattice QCD calculations.

\begin{figure}[h]
  \centering
  \includegraphics[width=0.5\textwidth,clip]{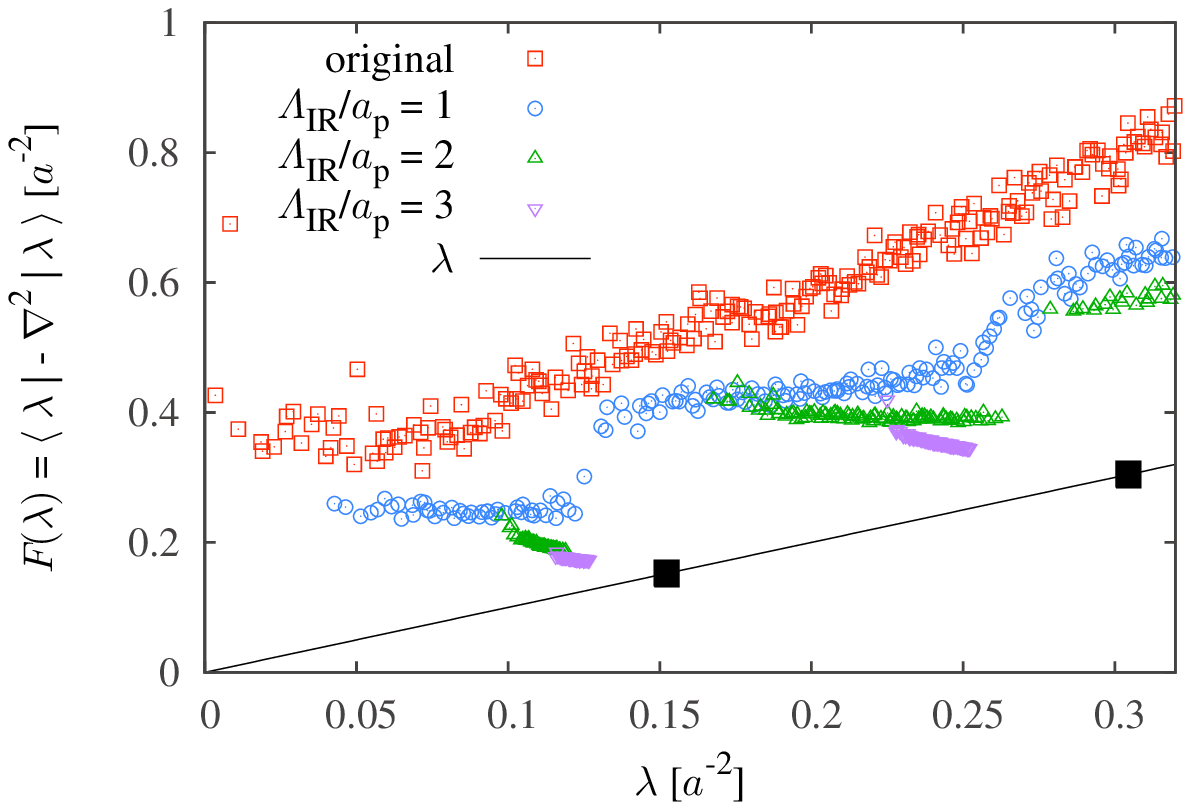}
  \includegraphics[width=0.5\textwidth,clip]{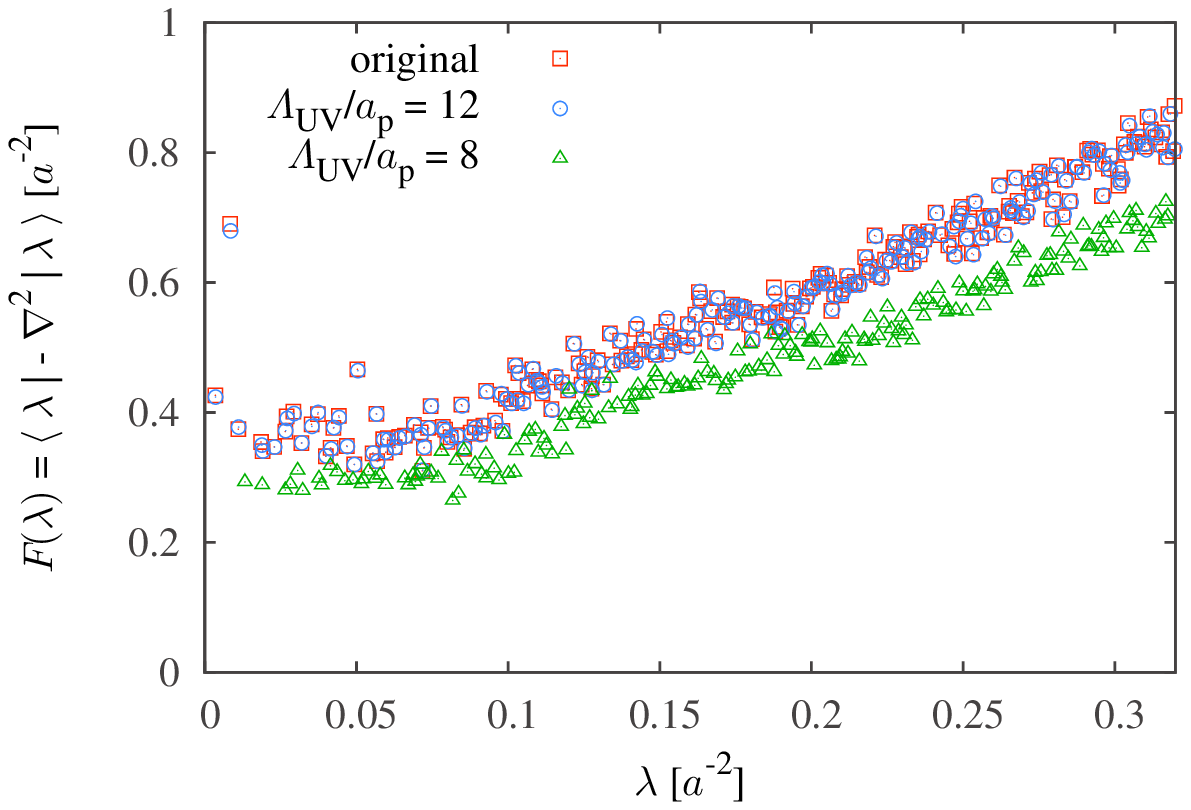}
  \caption{ \label{figFlambda}
    The scatter plot of the diagonal matrix element 
    $F(\lambda) \equiv \langle \lambda | - \nabla^2 | \lambda \rangle$
    with the IR/UV momentum cut,
    and original matrix element,
    for $16^4$ lattice with $a \simeq 0.15$fm, i.e., $a_p \equiv 2\pi/La \simeq 0.50$GeV.
    (a) IR-momentum cut with $\Lambda_{\rm IR}/a_p = 1,2$, and $3$.
    The solid-box symbols denote for non-zero free-field values,
    and the solid line for free-field value in continuum limit,
    i.e., $F(\lambda) = \lambda$.
    (b) UV-momentum cut with $\Lambda_{\rm UV}/a_p = 12$ and $8$.
  }
\end{figure}

  Figure \ref{figFlambda} is
  the scatter plot of the diagonal element $F(\lambda)$ with the IR/UV-momentum cut,
  for low-lying 250 eingenmodes.
  We also show the original $F(\lambda)$
  without momentum cut, and omit the trivial 8 zero-modes in this figure.
  In Fig.\ref{figFlambda}(a)
  the solid-box symbols denote the free-field values
  $\lambda = 4\sin^2(\pi/16)a^{-2} \simeq 0.15a^{-2}$
  and $\lambda = 8\sin^2(\pi/16)a^{-2} \simeq 0.30a^{-2}$
  in this $16^4$ lattice,
  and the solid line denotes free-field value $F(\lambda) = \lambda$ in the continuum theory.

  By the IR-momentum cut, near-zero modes of $F(\lambda)$ vanish 
  as in the IR-cut FP spectrum.
  $F(\lambda)$ changes into band-like structure from original smooth distribution,
  and tends to converge into free-field value on lattice.
  In Fig.\ref{figFlambda},
  one finds a ``flow'' of eigenmodes into free-field limit by the IR-momentum cut.
  Since both matrix element $F(\lambda)$ and FP spectrum $\rho(\lambda)$ converge 
  to free-field limit, the self-energy divergence condition (\ref{eqSelfEnergyDivergence})
  would not be satisfied by the IR-cut.
  Actually, the instantaneous potential becomes non-confining
  as shown in Fig.\ref{figColorCoulomb}(a).
  On the other hand, $F(\lambda)$ is almost unchanged 
  by the UV-momentum cut like the FP spectrum.

\subsubsection{Full FP eigenmodes}
  As shown above, the low-lying FP eigenmodes are largely 
  changed by the IR-momentum cut.
  Therefore, one may expect naively that low/high FP eigenmodes are 
  closely related to the IR/UV-momentum gluon, respectively.

  To investigate correspondence
  between gluonic momentum and the FP spectrum,
  it is meaningful to calculate full FP eigenmodes.
  However, it requires huge computational costs
  to perform full diagonalization of the FP operator for large volume lattices.
  Here, we adopt $8^4$ lattice with $\beta = 5.6$,
  which corresponds to the lattice spacing $a \simeq 0.25$fm
  and $a_p \simeq 0.62$GeV \cite{Suganuma:2011,Gongyo:2012}.
  The total number of the FP eigenmodes is $V_3 \times (N_c^2-1) = 8^3 \times 8 = 4096$.
  We evaluate the full FP eigenmodes using LAPACK \cite{LAPACK}.
  We have confirmed that the instantaneous potential and 
  momentum cut dependence in this $8^4$ lattice,
  and find that qualitatively the same results are obtained 
  in $16^4$ lattice with $\beta = 5.8$.
  Therefore, full-modes analysis is workable in this $8^4$ lattice.

\begin{figure*}
    \centering
    \includegraphics[width=0.49\textwidth]{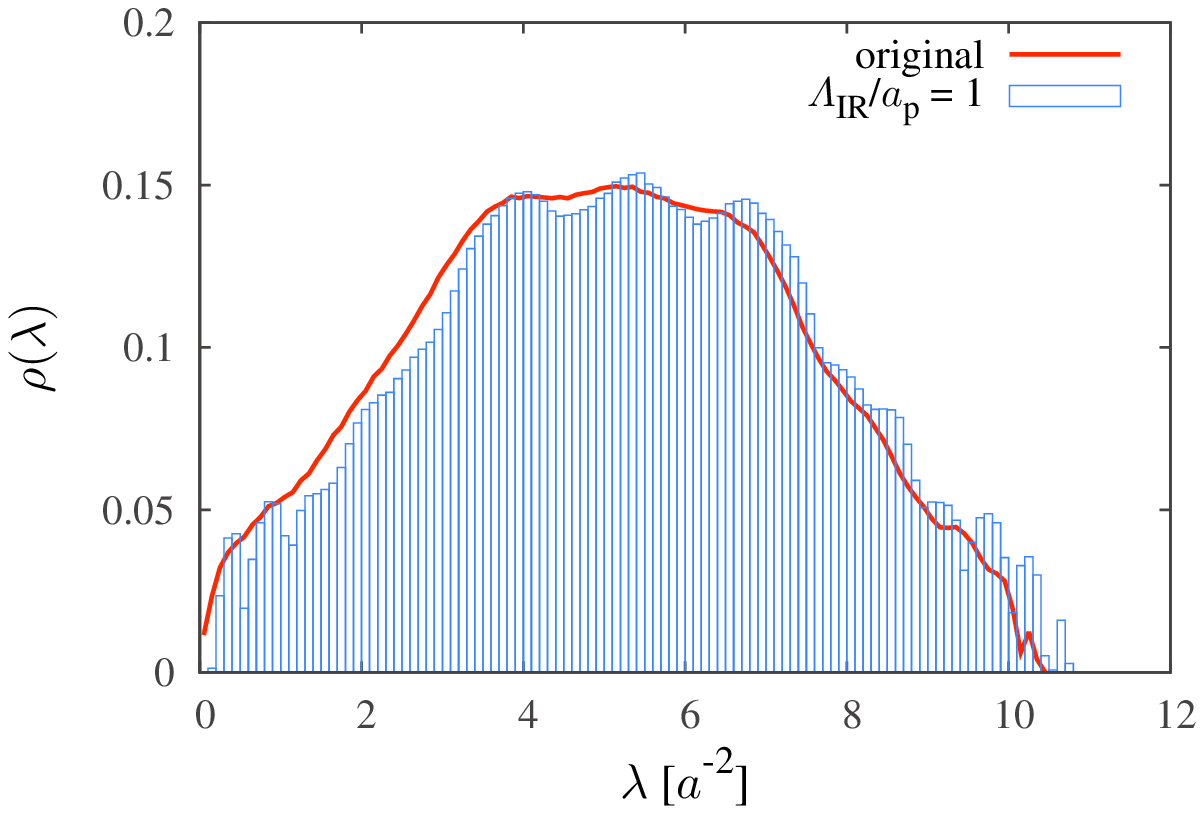}
    \includegraphics[width=0.49\textwidth]{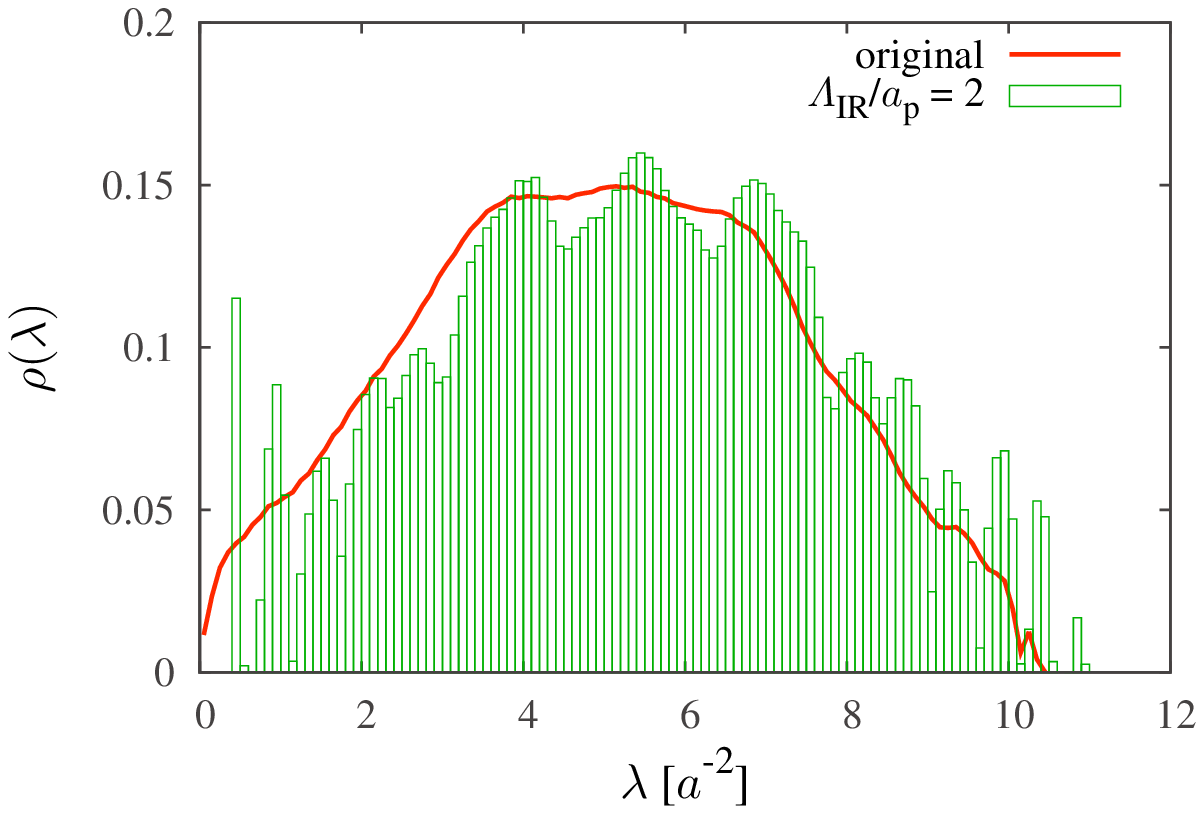}
    \includegraphics[width=0.49\textwidth]{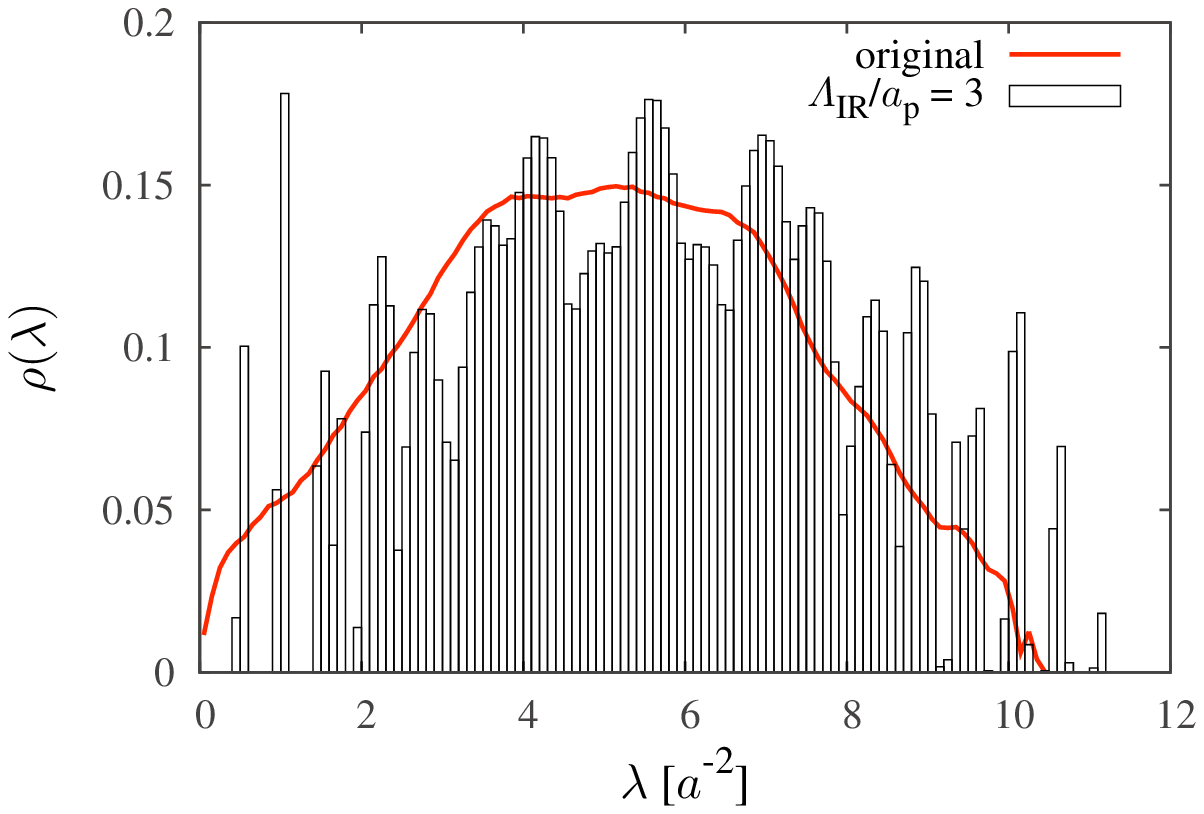}
    \includegraphics[width=0.49\textwidth]{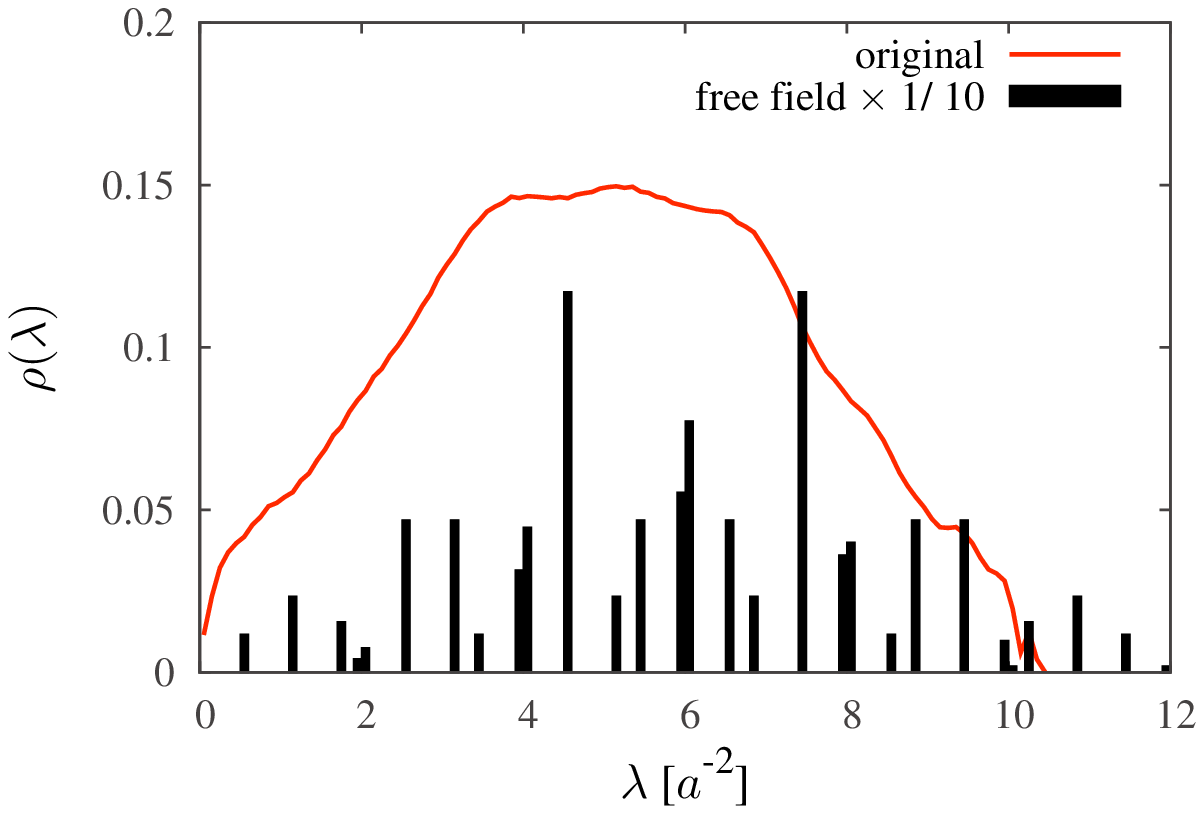}
    \caption{ \label{figFPspectrumIR}
      The full FP spectrum $\rho(\lambda)$ with the IR-momentum cut
      for $8^4$ lattice at $\beta = 5.6$, i.e.,
      $a \simeq 0.25$fm and $a_p \equiv 2\pi/La \simeq 0.62$GeV.
      The binwidth is taken as $\Delta \lambda = 0.1a^{-2}$.
      The solid curve denotes original spectrum.
      (a) IR-momentum cut with $\Lambda_{\rm IR}/a_p = 1$,
      (b) $\Lambda_{\rm IR}/a_p = 2$, (c) $\Lambda_{\rm IR}/a_p = 3$,
      (d) free field spectrum with rescaled by $1/10$. 
    }
  \end{figure*}

\begin{figure}
    \includegraphics[width=0.49\textwidth]{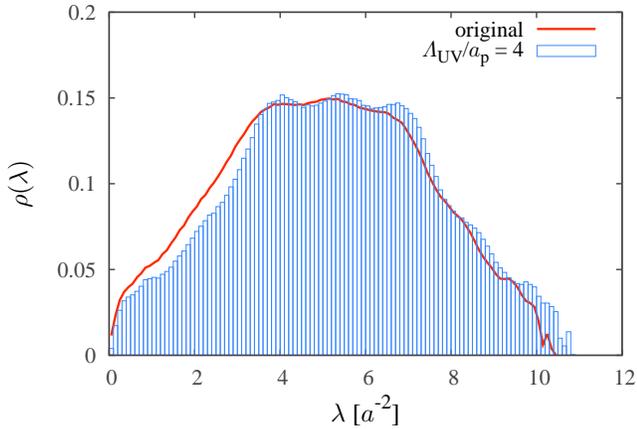}
    \caption{ \label{figFPspectrumUV}
      The full FP spectrum $\rho(\lambda)$ with the UV-momentum cut
      at $\Lambda_{\rm UV}/a_p = 4$ 
      for $8^4$ lattice at $\beta = 5.6$, i.e., 
      $a \simeq 0.25$fm and $a_p \equiv 2\pi/La \simeq 0.62$GeV.
      The binwidth is taken as $\Delta \lambda = 0.1a^{-2}$.
      The solid curve denotes original spectrum.
    }
  \end{figure}

  Figure \ref{figFPspectrumIR} and
  \ref{figFPspectrumUV} are the full FP spectrum 
  with the IR/UV momentum cut, respectively.
  The solid curve denotes original FP spectrum 
  in Figs.\ref{figFPspectrumIR} and \ref{figFPspectrumUV},
  and Fig.\ref{figFPspectrumIR}(d) is free-field spectrum in this lattice size.
  For the IR-cut, the FP spectrum is drastically changed
  from smooth one to multi-peak structure in the whole eigenvalue region.
  It is notable that 
  both low and high FP eigenmodes are affected by
  the low momentum gluon.
  The IR-cut spectrum clearly approaches to
  free-field spectrum, which is given by Eq.(\ref{eq:FPeigenFree}).
  In contrast, the FP spectrum is almost unchanged for UV-cut
  as shown in Fig.\ref{figFPspectrumUV}.

  Therefore, there are no direct correspondence
  between the IR/UV gluonic momentum and low/high FP eigenmodes, respectively.
  Figures \ref{figFPspectrumIR} and \ref{figFPspectrumUV} 
  indicate that the smooth structure of FP eigenmodes
  mainly originate from the low-momentum gluon.

  As a caution, 
  the UV-cut spectrum in Fig.\ref{figFPspectrumUV}
  seems to resemble the IR-cut one at $\Lambda_{\rm IR}/a_p = 1$
  as shown in Fig.\ref{figFPspectrumIR}(a).
  However, as we discussed the low-lying FP eigenmodes,
  the near-zero modes vanish in the IR-cut,
  which differs from the UV-cut spectrum.
  The near-zero eigenmodes are relevant for color-Coulomb energy,
  such a small difference of low-lying spectrum
  can affect on the confining property of the color-Coulomb energy.

\subsection{Color-Coulomb energy from FP eigenmodes}
  In Sec.III-A,
  we investigate the instantaneous potential instead of the color-Coulomb energy.
  In this subsection, 
  we directly analyze the color-Coulomb energy in Eq.(\ref{eq:VcoulFPmode})
  using the FP eigenmodes \cite{Nakagawa:2010}, 
  which are evaluated in Sec.III-B.

\begin{figure}[h]
  \centering
  \includegraphics[width=0.5\textwidth,clip]{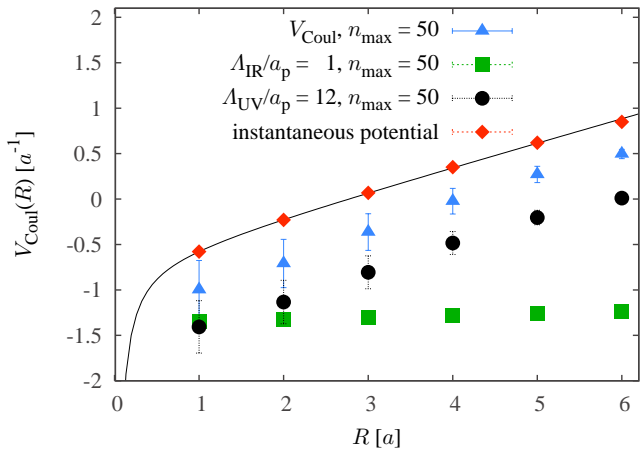}
  \caption{ \label{figFPspectrumSum}
  The color-Coulomb energy $V_{\rm Coul}(R)$ obtained with Eq.(\ref{eq:VcoulFPmode})
  using low-lying 50 FP eigenmodes apart from trivial zero-modes.
  For $16^4$ lattice at $\beta = 5.8$, i.e.,
  $a\simeq 0.15$fm and $a_p \equiv 2\pi/La \simeq 0.50$GeV,
  we show the IR-momentum cut with $\Lambda_{\rm IR}/a_p = 1$,
  the UV-momentum cut with $\Lambda_{\rm UV}/a_p = 12$,
  and original color-Coulomb energy.
  For comparison, the instantaneous potential is also added
  with the fitting curve of Coulomb plus linear form.
  An irrelevant constant is shifted.
  }
\end{figure}

  Figure \ref{figFPspectrumSum} shows
  the color-Coulomb energy with IR/UV-momentum cut
  using low-lying 50 FP eigenmodes apart from trivial 8 zero-modes.
  For comparison, we add the original (no momentum cut) color-Coulomb energy,
  and the instantaneous potential with the fitting curve of Coulomb plus linear form.
  Here, the color-Coulomb energy is calculated directly
  with Eq.(\ref{eq:VcoulFPmode}).

  As shown in Figs.\ref{figColorCoulomb} and \ref{figFPspectrumSum}, 
  the instantaneous potential and 
  the color-Coulomb energy 
  exhibit the similar momentum-cut dependence.
  The UV-cut color-Coulomb energy is almost the same as the original one,
  and the IR-cut energy becomes non-confining.

\subsection{Comparison with center-vortex removal}
  In this paper, 
  we have investigated the instantaneous potential
  and FP eigenmodes in terms of gluon-momentum components.
  In the context of the confinement mechanism, 
  there is an interesting similarity between 
  our results on the IR-momentum gluon cut 
  and those on the center-vortex removal \cite{Greensite:2003}, 
  and then we compare these two different operations 
  including their results in this subsection.

  The center vortex is an interesting object 
  appearing in the maximal center (MC) gauge
  \cite{DelDebbio:1997},
  and closely relates to the confinement \cite{Greensite:2003}
  and the infrared properties of QCD \cite{Quandt:2010,Chernodub:2011}.
  Actually, when the center vortex is removed from the QCD vacuum, 
  the string tension obtained from the Wilson loop vanishes 
  and the system becomes non-confining \cite{dFE99},
  which is observed in the IR-momentum cut of gluons \cite{Yamamoto:2008}.

  Also, as Greensite and Olejn\'ik pointed out, 
  by removing the center vortex in the MC gauge,
  the instantaneous potential
  in the Coulomb gauge becomes 
  non-confining \cite{Greensite:2003},
  which resembles the IR-cut of gluons, 
  as shown in Fig.\ref{figColorCoulomb}(a).
  Since the instantaneous potential
  is closely related to the color-Coulomb energy,
  the non-confining instantaneous potential
  reflects the change of the near-zero FP eigenmodes.
  In the center-vortex removal, 
  the FP spectrum becomes multi-peaks, 
  and matrix element $\langle \lambda | - \nabla^2 | \lambda \rangle$
  becomes band-like structure \cite{Greensite:2005}.
  These changes are quite similar to those of the IR-momentum cut of gluons, 
  as shown in Figs.\ref{figFPspectrumLow}(a), \ref{figFlambda}(a),
  and \ref{figFPspectrumIR}(a)-(c).
  In fact, both operations lead to the similar drastic change of FP eigenmodes. 
  In the center-vortex removal, however, 
  the FP spectrum does not coincide with the $\delta$-functional form 
  of the free-field case, but becomes multi-peaks with a finite width.
  Then, there still remains the significant difference between 
  ``no-vortex'' link-variables and free-field variables.

  In the gluon-momentum cut method,
  one can continuously change the link-variable to free-field one, 
  and accordingly, the FP spectrum continuously goes from a smooth function 
  to the $\delta$-functional form.
  In fact, unlike the center-vortex removal, our operation is continuous.
  In the momentum-cut method, according to the IR cut,
  we observe a ``flow'' of the matrix element $F(\lambda)$, 
  which continuously goes to the free-field limit, as shown 
  in Fig.\ref{figFlambda}(a).

  From the similarity between these two different operations, 
  it is also interesting to investigate 
  the correspondence between the center vortex in the MC gauge 
  and the gluon-momentum components in the Landau/Coulomb gauge.

\section{Summary and Discussion}
  In this paper, we have investigated the relation between
  the FP eigenmodes and gluon-momentum components in the Coulomb gauge
  using SU(3) lattice QCD calculations at the quenched level.
  The FP eigenmodes are considered to be important
  for confinement scenario in the Coulomb gauge.
  Especially, the low-lying eigenmodes
  lead the large color-Coulomb energy,
  and the color-Coulomb self-energy diverges
  from the enhancement of the near-zero eigenmode density.
  We have analyzed low-lying FP eigenmodes 
  with the IR/UV-momentum gluon cut,
  and also performed the full FP eigenmodes analysis.

  In the UV-momentum gluon cut, both color-Coulomb energy
  and FP eigenmodes are almost unchanged,
  which indicates that high-momentum gluons are irrelevant
  for confining scenario in the Coulomb gauge.
  In contrast, the color-Coulomb energy becomes non-confining,
  and the FP eigenmodes are drastically changed by the IR-momentum cut.
  In the IR-momentum gluon cut,
  the FP spectrum changes from original smooth one
  to the multi-peaks, which converges into free-field limit.
  We also note that the changes of the FP spectrum
  occurs in the whole eigenvalue region by low-momentum gluon cut.
  These results indicate the importance of the IR-momentum gluon
  for non-trivial structure of the FP eigenmodes.

  We comment on change of the FP eigenmodes in the continuum limit.
  By the IR-momentum cut, 
  the FP spectrum becomes the comb-like shapes, 
  since the spectrum converges into the discrete free-field one on lattice.
  However, in the continuum limit, 
  the FP spectrum changes into smooth free-field form
  $\rho(\lambda) \sim \lambda^{1/2}$ 
  from near-zero enhanced spectrum in QCD.
  Therefore, by the IR-momentum cut,
  the modification of the spectrum would be smaller in the continuum theory
  than the discrete case as shown in Fig.\ref{figFPspectrumIR}.

  In this paper, we have investigated the FP eigenmodes
  in terms of the gluon-momentum components.
  It is also interesting to investigate QCD properties
  in terms of the FP eigenmode,
  since the FP eigenmodes are considered to be important for confinement
  in the Coulomb gauge.

\begin{acknowledgments}
  T.I. is supported by a Grant-in-Aid for JSPS Fellows [No. 23-752], 
  and H.S. is supported in part by the Grant for Scientific
  Research [(C) No.23540306, Priority Areas ``New Hadrons'' (E01:21105006)] 
  from the Ministry of Education, Culture, Science and Technology (MEXT) of Japan.
  This work is supported by the Global COE Program,
  ``The Next Generation of Physics, Spun from Universality and Emergence".
  The lattice QCD calculations have been done on NEC-SX8 and NEC-SX9 at Osaka University.
\end{acknowledgments}

\end{document}